\documentclass[sigconf,nonacm]{acmart}

\usepackage{newfloat}
\usepackage{makecell}
\usepackage{listings}
\usepackage{amsmath}
\usepackage{epigraph}
\usepackage{booktabs}
\usepackage{multirow}
\usepackage{graphicx}
\usepackage{caption}
\usepackage{adjustbox}
\usepackage[most]{tcolorbox}
\usepackage{CJKutf8}
\usepackage{enumitem}
\usepackage{color}
\usepackage{natbib}

\usepackage{pifont}
\usepackage{balance}

\newcommand{\ie}{\textit{i}.\textit{e}.}
\newcommand{\eg}{\textit{e}.\textit{g}.} 

\usepackage[table]{xcolor}
\usepackage{xspace}

\definecolor{mycolor}{RGB}{134,150,167}
\newtcbtheorem[auto counter, number within = section]{cmt}{}{
	colbacktitle = mycolor, colframe = mycolor,
	colback = mycolor!10!white,
	fonttitle=\bfseries,
 top = 1pt,          
  bottom = 1pt,       
  left = 4pt,         
  right = 4pt,        
  boxsep = 2pt,       
}{t}

\AtBeginDocument{%
  }
\setcopyright{none}
\begin{document}
\begin{CJK}{UTF8}{gbsn}
\newcommand{\ourmodel}{MR.Rec\xspace}

\author{Jiani Huang*}
\email{jianihuang01@gmail.com
}
\affiliation{%
  \institution{The Hong Kong Polytechnic University}
  \city{Hong Kong SAR, China}
  \country{}
}

\author{Xingchen Zou*}
\email{xzou428@connect.hkust-gz.edu.cn}
\affiliation{%
  \institution{ The Hong Kong University of Science and Technology (Guangzhou)}
  \city{Guangzhou, China}
  \country{}
}

\author{Lianghao Xia\dag}
\email{aka_xia@foxmail.com}
\affiliation{%
  \institution{The University of Hong Kong}
  \city{Hong Kong SAR, China}
  \country{}
}

\author{Qing Li\dag}
\email{qing-prof.li@polyu.edu.hk}
\affiliation{%
  \institution{The Hong Kong Polytechnic University}
  \city{Hong Kong SAR, China}
  \country{}
}

\thanks{*These authors contributed equally to this work.\\ \dag Corresponding author: Lianghao Xia, Qing Li.}

\title{MR.Rec: Synergizing \underline{M}emory and \underline{R}easoning for Personalized \underline{Rec}ommendation Assistant with LLMs}

\begin{abstract}

The application of Large Language Models (LLMs) in recommender systems faces key challenges in delivering deep personalization and intelligent reasoning, especially for interactive scenarios. Current methods are often constrained by limited context windows and single-turn reasoning, hindering their ability to capture dynamic user preferences and proactively reason over recommendation contexts. To address these limitations, we propose MR.Rec, a novel framework that synergizes memory and reasoning for LLM-based recommendations. To achieve personalization, we develop a comprehensive Retrieval-Augmented
Generation (RAG) system that efficiently indexes and retrieves relevant external memory to enhance LLM personalization capabilities. Furthermore, to enable the synergy between memory and reasoning, our RAG system goes beyond conventional query-based retrieval by integrating reasoning-enhanced
memory retrieval. Finally, we design a reinforcement learning framework that trains the LLM to autonomously learn effective strategies for both memory utilization and reasoning refinement. By combining dynamic memory retrieval with adaptive reasoning, this approach ensures more accurate, context-aware, and highly personalized recommendations.
Extensive experiments demonstrate that MR.Rec significantly outperforms state-of-the-art baselines across multiple metrics, validating its efficacy in delivering intelligent and personalized recommendations. 
We will release code and data upon paper notification.
\end{abstract}


\maketitle

\section{Introduction}
\label{sec:intro}

As digital content continues to grow, recommender systems (RSs) have become widely adopted across web platforms to filter massive information spaces and provide users with personalized content~\cite{zhang2019deep}. With the rapid advancement of artificial intelligence, user expectations have evolved toward more intelligent and personalized recommendation assistants. Specifically, users now expect these systems to understand natural language queries—similar to voice assistants like Siri—and provide immediate, personalized recommendations~\cite{huang2025towards,zhang2025survey,han2025rethinking}. Traditional RSs, such as matrix factorization (MF)~\cite{mnih2007probabilistic} or graph neural network (GNN)–based methods~\cite{wu2022graph}, primarily rely on implicit signals from interaction histories (e.g., clicks, purchases). While effective in many contexts, they cannot interpret natural language inputs, limiting their ability to function as intelligent, conversational recommendation assistants.

\begin{figure}[t!]
\centering
\includegraphics[width=1.0\linewidth]{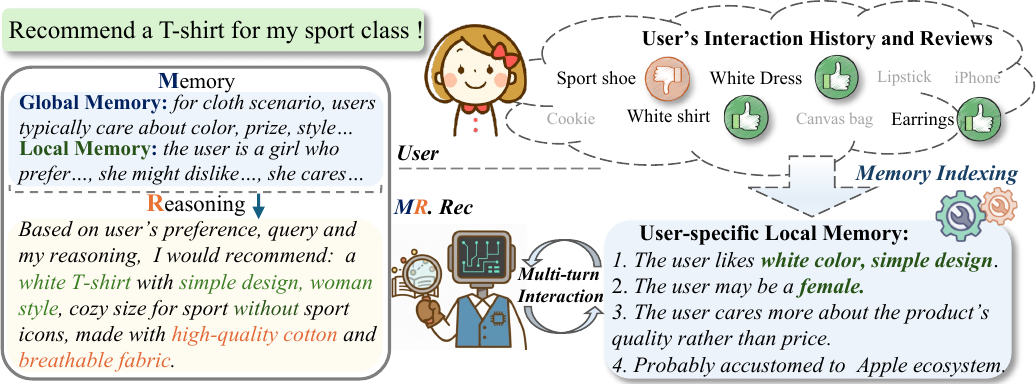}
\vspace{-1.8em}
\caption{Illustration of Personalized and Intelligent LLM-based Recommendation Assistant.}
\label{fig:intro}
\vspace{-1.5em}
\end{figure}

The advent of large language models (LLMs) has opened new possibilities for developing intelligent, interactive recommendation assistants in various domains~\cite{zhao2024recommender,wang2025knowledge}. 
With their advanced language understanding capability, extensive general knowledge and instruction-following ability~\cite{zhou2023instruction}, LLMs have the potential to understand user natural language queries and serve as intelligent recommendation assistants. Therefore, existing research has explored the integration of LLMs into recommendation tasks~\cite{fang2024multi,he2025reindex}. Despite recent advancements, challenges remain in achieving truly personalized and intelligent recommendation assistants, primarily due to the limited ability of LLMs to memorize user preferences and reason effectively on recommendation tasks.

One of the primary obstacles lies in \textbf{personalization}, as the assistant must be able to maintain a comprehensive \underline{\textit{user memory}} that captures user profiles, evolving preferences, and contextual histories~\cite{zhang2024mindmemory,magister2024way}. This persistent memory allows the assistant to anticipate user needs and deliver relevant recommendations seamlessly. For example, as illustrated in Figure~\ref{fig:intro}, when a user asks for T-shirt recommendations, the pre-modeled user memory enables the assistant to understand the user's past preferences (e.g., liking white color, simple design, and high quality) and inferred user profile (e.g., potential gender), thereby facilitating highly personalized and contextual recommendations that go beyond simple query matching. Existing LLM-based RSs typically inject user personalization information into LLMs either by prompting recent interaction histories~\cite{lyu2023llm,wang2025tunable} or by providing pre-generated, static user profiles~\cite{wang2025lettingo,zhou2024language}. However, these methods are often constrained by the limited context window of LLMs, which prevents the assistant from fully capturing a user’s profile and evolving preferences. In addition, they may introduce irrelevant or noisy information that is not aligned with the current query, thereby distracting the model and reducing its ability to generate accurate recommendations.

While personalization is essential, it does not fully address the challenge of building a capable recommendation assistant. A complementary requirement is \textbf{intelligence}, namely the ability to actively explore the factors behind effective recommendations and perform in-depth multi-step \underline{\textit{reasoning}}~\cite{zhang2024llm,ferrag2025llm}. In the recommendation domain, such reasoning is essential for interpreting user queries, uncovering underlying intents, and generating accurate suggestions. As illustrated in Figure~\ref{fig:intro}, when a user requests a T-shirt for sports, the recommendation assistant should reason that breathable fabric is essential for athletic activities, and then integrate this reasoning with the user’s memory to generate a more intelligent and contextually appropriate recommendation. Recent works have begun to activate LLMs' reasoning in recommendations~\cite{bismay2024reasoningrec,wang2024llmrg}; for instance, Tang et al.~\cite{tang2025think} enhance user representations through implicit multi-step reasoning in sequential recommendation. Nevertheless, these efforts generally constrain the LLM to reason only over the limited information provided in the pre-defined prompt template, without enabling it to actively explore which additional user memories are helpful for resolving the current recommendation problem. This decoupled paradigm between reasoning and memory restricts the assistant to shallow reasoning, limiting its capacity to effectively leverage stored memory and perform coherent, multi-step reasoning essential for generating personalized and contextually grounded recommendations.

To address these challenges, we propose \textbf{\ourmodel}, a model that synergizes \textbf{M}emory and \textbf{R}easoning for LLM-based \textbf{Rec}ommendation assistants. \textbf{First}, we develop a comprehensive Retrieval-Augmented Generation (RAG) system that efficiently indexes and retrieves relevant external memory to enhance LLM personalization capabilities. Our RAG system employs a hierarchical memory indexing mechanism that consists of \underline{\textit{User-specific Local Memory}} and \underline{\textit{Cross-user Global Memory}}. User-specific Local Memory captures user-specific information at multiple granularities, progressively organizing interaction histories, preference patterns, and user profiles to comprehensively represent personalized preferences, while Cross-user Global Memory aggregates cross-user knowledge to provide broader experiential understanding for diverse recommendation scenarios. \textbf{Second}, to enable the synergy between memory and reasoning, our RAG system goes beyond conventional query-based retrieval by integrating reasoning-enhanced memory retrieval. Rather than directly retrieving based on surface-level query similarity, our approach first leverages reasoning to identify implicit user preferences and relevant preference dimensions, then selectively retrieves memory segments accordingly. Furthermore, we extend LLM reasoning beyond simple chain-of-thought to encompass multi-step memory retrieval stages, creating an adaptive framework where reasoning and information gathering are interleaved to enable dynamic memory exploration and evidence-grounded inference that significantly enhances reasoning depth and context awareness. \textbf{Finally}, we design a reinforcement learning framework that trains the LLM to autonomously learn effective strategies for both memory utilization and reasoning refinement. Through diverse reward mechanisms that guide memory retrieval, reasoning quality, and recommendation accuracy, our end-to-end optimization enables the model to develop adaptive behaviors that optimize long-term recommendation performance while efficiently leveraging the hierarchical memory structure.

In summary, our contributions can be summarized as follows: 
\begin{itemize}[leftmargin=*,itemsep=0pt, parsep=0pt, topsep=0pt, partopsep=0pt]
    \item We propose a comprehensive RAG system with hierarchical memory indexing that efficiently captures and retrieves personalized and cross-user knowledge, significantly extending external memory capabilities for LLM-based recommendations.\\\vspace{-0.12in}
    \item We develop a reinforcement learning paradigm with novel rewards to harmonize memory access and recommendation reasoning, enabling the LLM to autonomously learn effective memory utilization strategies for final recommendation.\\\vspace{-0.12in}
    \item Extensive experiments demonstrate that our \ourmodel outperforms state-of-the-art baselines, validating its efficacy in delivering intelligent, personalized recommendations.
\end{itemize}

\section{Preliminary}
\label{sec:preliminary}

\noindent\textbf{Memory-Enhanced Recommendation}.
Memory enables recommenders to retain user interaction histories and contextual information for personalized recommendations~\cite{wu2025human}. Given users $\mathcal{U} = \{u_1, u_2, \dots, u_{|U|}\}$ and items $\mathcal{I} = \{i_1, i_2, \dots, i_{|I|}\}$, each user $u$ has interaction history $\mathcal{H}_u = \{(i_t, r_t)\}_{t=1}^{T_u}$, where $i_t$ is the interacted item and $r_t$ is auxiliary signals (ratings, reviews). A memory function $\mathcal{M}(\cdot)$ transforms this history into structured memory:
\abovedisplayskip=0pt
\belowdisplayskip=0pt
\begin{align}
\mathcal{M}_u = \mathcal{M}(\mathcal{H}_u),
\end{align}
where $\mathcal{M}_u$ encodes preferences as embeddings, textual summaries, or hybrid structures for LLM retrieval during recommendation.\vspace{0.05in}

\noindent\textbf{Reasoning in LLM-based Recommendation}.
Effective LLM-based recommendation requires explicit reasoning to decompose complex queries and adaptively utilize relevant memory. Rather than direct prediction, LLMs should generate interpretable reasoning trajectories $\mathcal{T}$ that systematically trace from query understanding through memory analysis to final recommendation justification. This reasoning-driven approach enables dynamic memory exploration and provides transparent recommendation decisions.\vspace{0.05in}

\noindent\textbf{Problem Statement}.
We develop a memory-enhanced recommendation system that generates personalized recommendations through explicit reasoning. Given user $u$ with textual query $q_u$ and memory $\mathcal{M}_u$, our system generates an ideal item profile $i^*$ via reasoning-enhanced LLM:
\begin{align}
i^* = \text{LLM}(\mathcal{P}(q_u, \mathcal{M}_u); \mathbf{\Theta}),
\end{align}
where $\mathcal{P}(\cdot)$ integrates query and memory into structured prompts, and $\mathbf{\Theta}$ are learnable parameters optimized for reasoning quality. We then retrieve top-$k$ items from item pool $\mathcal{I}$ based on similarity:
\begin{align}
R_u = \text{arg top-}k_{i \in \mathcal{I}} \text{sim}(i^*, \text{metadata}(i)).
\end{align}
Our objective is to jointly optimize memory utilization and reasoning parameters $\mathbf{\Theta}$ to improve recommendation accuracy.

\section{Methodology}
\label{sec:method}
\begin{figure*}
    \centering
    \vspace{-0.1in}
    \includegraphics[width=1.0\linewidth]{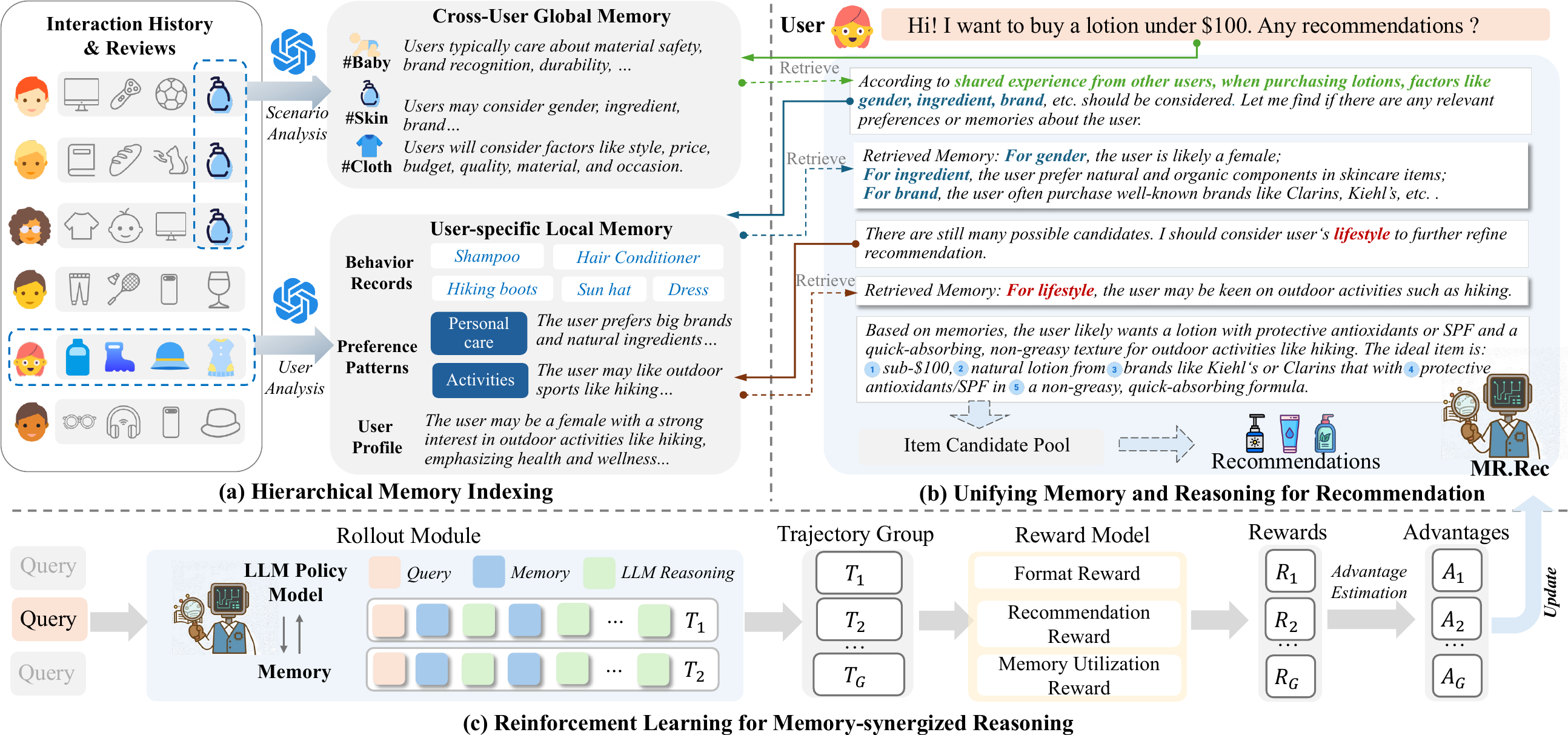}
   \vspace{-1.8em}
    \caption{The overall model architecture of the proposed \ourmodel\ framework. }
    \label{fig:method}
    \vspace{-1em}
\end{figure*}

This section elaborates technical details of our \ourmodel\ framework. The overall framework is depicted in Figure~\ref{fig:method}.

\subsection{Memory Indexing for RAG Recommender}
To enable LLMs to effectively and efficiently obtain external information from vast user interaction records, we propose a retrieval-augmented generation \textbf{(RAG)-based External Memory Mechanism} for recommendation. This RAG module adopts a \textbf{Hierarchical Memory Indexing} method. Specifically, effective memory indexing for recommendation faces two key challenges. First, using raw interaction histories can exceed context limits and cause overfitting to recent behavior while introducing irrelevant noise. Second, recommender systems need both fine-grained user-specific memories and comprehensive knowledge aggregated across users to provide robust recommendations. Therefore, we compress interaction data into the following two complementary memory structures:

\subsubsection{\bf User-specific Local Memory}
User-specific memory involves users' purchase histories, ratings, and reviews~\cite{pradel2011case}. This data can easily exceed LLMs' context window, but using users' recent interactions may overlook their long-term preferences and overfit their short-term behaviors. This makes it infeasible to directly use the raw interaction data. More essentially, raw interaction histories are inherently noisy: not all past behaviors are equally relevant to the current recommendation query. For instance, past grocery purchases are irrelevant when recommending home appliances, while electronics-related behaviors provide more reliable signals.

To address these challenges, we organize user preferences at multiple granularities, enabling selective retrieval of relevant behavioral patterns while filtering noise, specifically including:
\begin{itemize}[leftmargin=*,itemsep=0pt, parsep=0pt, topsep=0pt, partopsep=0pt]
\item \textbf{Behavior Records}: It preserves the original interaction data of a user, such as purchased items, ratings, and reviews.\\\vspace{-0.12in}

\item \textbf{Preference Patterns}: Moving beyond raw behavioral records, this part indexes preference patterns within item categories. We partition interaction history into different domains (\eg, electronics, clothing) and apply LLM to summarize compact preference descriptions. This filters out irrelevant cross-category signals while preserving salient information. Formally, given user interaction history $H_u$, we partition it into subsets ${H_u^c}$ by category $c$ and use LLM to summarize category-specific user preference $P_u^c$:
\begin{align}
    P_u^c = f_{\mathrm{LLM}}(H_u^c).
\end{align}

\item \textbf{User Profile}: 
To model coherent preference structures across domains, we construct a higher-level abstraction integrating fragmented category-specific patterns into consistent cross-category preference structures. Formally, the user profile is expressed as:
\begin{align}
    U_u = f_{\mathrm{LLM}}\!\left(\{P_u^c : c \in \mathcal{C}_u\} \right),
\end{align}
\belowdisplayskip=0pt
where $\mathcal{C}_u$ denotes the set of categories associated with user $u$. The resulting profile $U_u$ serves as a compact, high-level summary that complements fine-grained preference patterns.
\end{itemize}
Since preference patterns and user profiles are textual summaries, we apply chunking to decompose them into smaller, semantically coherent segments, enabling efficient storage and selective retrieval of the most pertinent information during recommendation.

\subsubsection{\bf Cross-User Global Memory}
To capture shared behavioral patterns from the entire dataset, \ourmodel constructs a global memory that captures common decision-making dimensions across users within similar recommendation scenarios, serving as a complementary knowledge source. For example, in baby product recommendations, such memory reveals critical domain-specific factors like safety certifications, ergonomic design, and age-appropriate features that users typically consider but rarely explicitly mention.

Concretely, for each recommendation scenario $s$ (\eg, a category in e-commerce), we sample a set of queries $Q_s = \{q_1, q_2, \dots, q_m\}$ from the training corpus, each paired with its ground-truth item $i^+ \in \mathcal{I}_s$. To uncover latent dimensions of decision-making, we incorporate negative items $\mathcal{I}^-_s \subset \mathcal{I}_s$ from the same scenario that were not chosen. These alternatives offer valuable contrasts, enabling the model to identify why the selected item was favored and to derive more granular, adaptable decision criteria. Formally, we define the global memory construction process as follows:
\begin{align}
    M_{global} = f_{\mathrm{LLM}}\big( \{ (q, i^+, \mathcal{I}^-_s) \mid q \in Q_s \} \big),
\end{align}\belowdisplayskip=0pt
where $M_s$ denotes the global memory for scenario $s$, and $f_{\mathrm{LLM}}(\cdot)$ extracts organized aspects and rationales.

\subsection{Unifying Memory and Reasoning for Recommendation}

\subsubsection{\bf Reasoning-enhanced Memory Retrieval}
With the indexed recommendation memory, our RAG-based recommender retrieves user interaction patterns relevant to the query to enhance recommendation accuracy. A critical challenge in memory retrieval for recommendations is \textbf{implicit user preferences} that remain unstated in queries.
Different recommendation scenarios engage diverse preference dimensions that users often omit unintentionally.
For instance, when requesting clothing recommendations, users may have unstated considerations about material, style, or brand reputation that are crucial for personalized suggestions.
This challenge causes existing static retrieval methods to fail, as they rely solely on surface-level query similarity without understanding the deeper preference dimensions. Consequently, such approaches can incorporate irrelevant data or miss critical information, limiting the assistant's ability to provide personalized recommendations.

To address these limitations, we propose reasoning-enhanced memory retrieval, which integrates reasoning and memory access in a dynamic, iterative process. Given a user query, the LLM first analyzes the recommendation scenario to identify relevant preference aspects that should guide personalization, leveraging global memory to understand typical evaluation criteria. The LLM then uses these reasoning-inferred aspects to selectively retrieve segments from user-specific local memory, enabling progressive refinement of understanding. Formally, we identify preference aspects by:
\abovedisplayskip=0pt
\begin{align}
    \mathcal{A}_q = f_{\mathrm{LLM}}(q, M_{\text{global}}),
\end{align}
where $f_{\mathrm{LLM}}(\cdot)$ represents the reasoning process over the query and global memory, and $\mathcal{A}_q$ denotes the identified preference aspects.

We then use the identified aspects to selectively retrieve relevant segments from the user-specific local memory, reducing noise and enhancing personalization:
\begin{align}
    \hat{M}_u(q) = g_{\mathrm{retrieval}}(\mathcal{A}_q, M_{\text{local}}),
\end{align}
where $g_{\mathrm{retrieval}}(\cdot)$ denotes the LLM-based retrieval function conditioned on the reasoning-inferred aspects, and $\hat{M}_u(q)$ represents the retrieved local memory segments.

\vspace{-0.1in}
\subsubsection{\bf Retrieval-Augmented Item Generation}
Finally, the LLM integrates the retrieved memories to generate an ideal item profile that aligns with both the user's query and personalized preferences:
\begin{align}
    \mathcal{I}_u(q) = f_{\mathrm{LLM}}\big(q, \mathcal{A}_q, \hat{M}_u(q)\big),
\end{align}
where $\mathcal{I}_u(q)$ represents the ideal item profile that captures the optimal item characteristics for the current query.

Through this reasoning-enhanced retrieval mechanism, the assistant dynamically leverages both global and local memory to produce contextually aware and highly personalized recommendations, treating memory as an active resource that evolves with the reasoning process rather than static input.

\vspace{-0.1in}
\subsubsection{\bf Enhancing Reasoning with Memory Retrieval}
Except that our memory retrieval method gains enhancement from LLM reasoning, we highlight that this method also represents an important enhancement over existing reasoning methods for recommendation tasks. Effective reasoning is crucial for understanding complex user preferences and making contextually appropriate connections between user needs and available options. Current recommendation LLMs typically employ chain-of-thought reasoning or prompt-based inference on static input, but these approaches suffer from limited reasoning depth as they cannot dynamically explore relevant information during the inference process, leading to suboptimal personalization and reduced context awareness. Our method enhances LLM reasoning by extending it to encompass memory retrieval stages, creating an adaptive framework where reasoning and information gathering are interleaved.

Specifically, the LLM iteratively analyzes the recommendation scenario, identifies relevant preference dimensions, and retrieves corresponding memory entries to refine its understanding. This enables the LLM to treat memory as a dynamic resource aligned with the evolving reasoning process, significantly enhancing reasoning capabilities through multi-step, evidence-grounded inference that adapts to task complexity. Importantly, this extended reasoning mechanism can be jointly optimized through our reinforcement learning framework, allowing the model to learn optimal strategies for both memory exploration and reasoning refinement.

\vspace{-0.08in}
\subsection{Reinforcement Learning for Memory-synergized Reasoning}
To enhance our memory-synergized reasoning, we fine-tune the base LLM through reinforcement learning. Our \ourmodel requires robust reasoning and effective memory exploration through multi-turn interactions and iterative retrieval. However, comprehensive annotation for such complex processes is prohibitively resource-intensive, making supervised fine-tuning impractical.

To overcome this, we adopt a multi-turn reinforcement learning framework. This approach allows the LLM to actively explore the recommendation and memory environment, refining its policy through iterative interactions. By leveraging feedback signals instead of static annotations, the model develops adaptive behaviors that optimize long-term recommendation accuracy and efficient memory utilization. Specifically, inspired by Group Relative Policy Optimization (GRPO)~\cite{guo2025deepseek}, we adopt a reinforcement learning with verifiable rewards framework that incorporates multi-turn feedback signals into policy optimization. For a given query $q$, the assistant is prompted multiple times to analyze the query, retrieve potentially useful memory entries, and reason through these before generating $G$ final candidate responses $\{o_1, o_2, \dots, o_G\}$. A reward model then assigns reward scores $\{r_1, r_2, \dots, r_G\}$ to each candidate response. Based on these scores, relative advantages $A(o_i)$ are computed, which indicate the comparative quality of different outputs and guide subsequent policy optimization.

To guide the model toward the desired behavior, we design a set of multi-faceted reward functions:
\vspace{-0.07in}
\begin{itemize}[leftmargin=*]
  \item \textbf{Format Reward} (\(R_{\text{format}}\)). To extract the desired answers from the reasoning process, we require the LLM to output its final answer in a prescribed format. The reward is assigned as 1 if the format is correct, and 0 otherwise.  

  \item \textbf{Recommendation Reward} (\(R_{\text{rec}}\)). To quantify recommendation quality, we define a reward $R_{\text{rec}} = \text{nDCG}@1000 + \text{nDCG}@100$ where  nDCG@1000 mitigates reward sparsity, and nDCG@100 emphasizes fine-grained ranking among top candidates.

  \item \textbf{Memory Utilization Reward} (\(R_{\text{mem}}\)). To encourage reasoning grounded in memory, the LLM is required to perform a memory retrieval step. The reward is defined as a binary indicator, where the reward is 1 if the model successfully calls memory retrieval during generation, and 0 otherwise.
\end{itemize}
\vspace{-0.05in}
The overall reward model is a weighted combination of the three reward components, with weights $w_1, w_2, w_3$:
\begin{align}
    r = w_1 R_{\text{format}} + w_2 R_{\text{rec}} + w_3 R_{\text{mem}}.
\end{align}

Finally, the optimization objective is formulated as a clipped policy gradient similar to PPO:
\begin{equation}
        \begin{aligned}
        & \mathcal{J}(\theta) = \mathbb{E}_{(q,gt) \sim \mathcal{D}, \{o_i\}_{i=1}^G \sim \pi_{\theta_{\text{old}}}(\cdot \mid q)} = &  \\
        & \left[ \frac{1}{N} \sum_{i=1}^G \sum_{t=1}^{|o_i|} \min \left( \frac{\pi_\theta(o|q)}{\pi_{\theta_{\text{old}}}(o|q)} A_{i,t},\text{clip} \left(\frac{\pi_\theta(o|q)}{\pi_{\theta_{\text{old}}}(o|q)}, 1-\epsilon, 1+\epsilon\right) A_{i,t} \right) \right],
        \label{eq:obj}            
        \end{aligned}
    \end{equation}
where $A_{i,t}$ denotes the advantage of each token derived from relative reward signals, which equals $A(o_i) = \frac{r_i - \operatorname{mean}(\textbf{r})}{\operatorname{std}(\textbf{r})}$. Importantly, during optimization, we mask the retrieved memory tokens to ensure that the advantage estimation depends only on the assistant’s reasoning and recommendation outputs.
\vspace{-0.05in}
\section{Experiments}
\label{sec:experiments}
In this section, we evaluate the proposed \ourmodel\ framework to answer the following research questions (RQs):
\begin{itemize}[leftmargin=*,itemsep=0pt, parsep=0pt, topsep=0pt, partopsep=0pt]
    \item \textbf{RQ1:} How does \ourmodel perform in providing personalized recommendation compared to baseline methods?
    \item \textbf{RQ2:} What is the contribution of each main component and different types of memory to the overall performance of \ourmodel?
    \item \textbf{RQ3:} Does our RAG-based memory mechanism retrieve genuinely beneficial information to address recommendation queries?
    \item \textbf{RQ4:} How is the efficiency of indexing and retrieval of our \ourmodel?
    \item \textbf{RQ5:} How do different model settings impact the performance of \ourmodel, including hyperparameter settings, LLM backbones, and different retriever configurations?
\end{itemize}

\vspace{-0.05in}
\subsection{Experimental Setup}
\subsubsection{\bf Datasets}
We construct our dataset based on the Amazon-C4~\cite{hou2024bridging} dataset, which provides user queries generated by ChatGPT from product reviews. These review-based queries are often highly detailed, implicitly covering nearly all aspects of a user’s preferences. However, such exhaustive queries are not representative of typical user behavior in real-world recommendation scenarios. To better evaluate our method and compare it with different baselines in leveraging user memory for personalized recommendations, we simplify these queries using GPT-o3-mini to remove some of the detailed preference information. More dataset statistics and the prompt used for query simplification can be found in \textbf{Appendix~\ref{app:dataset}}.

\vspace{-0.05in}
\subsubsection{\bf Baselines}
We evaluate our method against a set of representative models, which are divided into two groups. The first group consists of general-purpose LLM backbones: \texttt{GPT-4o}, \texttt{DeepSeek-R1}, and \texttt{Qwen-2.5-3B-Instruct}. The second group comprises models specifically fine-tuned for recommendation tasks: \texttt{BLAIR}~\cite{hou2024bridging}, a sentence embedding model pretrained on user review and item metadata pairs using a contrastive objective, and \texttt{Rec-R1}~\cite{lin2025rec}, which directly optimizes LLM generation via feedback from a fixed, black-box recommendation model. To investigate both the impact of memory and the models’ ability to leverage user memory for recommendation, we consider three memory settings for these models:

\begin{itemize}[leftmargin=*,itemsep=0pt, parsep=0pt, topsep=0pt, partopsep=0pt]
    \item \textbf{w/o Memory (Query-only):} Models receive only the current user query, without access to any historical interactions.
    \item \textbf{w/ Naive Memory (Query + user interaction history):} Except queries, models are provided with the latest interaction history of the user, representing straightforward memory integration.
    \item \textbf{w/ Static Memory (Query + pre-generated user summary):} Models are provided with a pre-constructed summary of the user’s preferences, representing static summarized memory.
\end{itemize}

\vspace{-0.05in}
\subsubsection{\bf Evaluation settings}
For each model under the corresponding memory setting, given a user query, we require the model to generate either the ideal item profile or its embedding, which is then used to retrieve items from the candidate pool. Recommendation performance is evaluated using standard metrics, including nDCG@100, nDCG@1000, Recall@100 and Recall@1000.

\vspace{-0.05in}
\subsubsection{\bf Implementation details.} We select Qwen-2.5-3B-Instruct as the backbone LLM in our method. During training, we use a learning rate
of 1e-6, a group size of 5, and set the maximum response length
to 768. Training is conducted for up to 5 epochs with early stopping (patience = 1). In our method, the hyperparameters for weighted reward are set \textbf{as $w_1=0.1, w_2=5, w_3=0.1$}. More implementation details can be found in \textbf{Appendix~\ref{app:imp}}.

\vspace{-0.05in}
\subsection{Overall Performance (RQ1)}

\begin{table*}[t]
  \centering
  \vspace{-0.1in}
  \caption{The overall performance of baselines and \ourmodel. The bold/underline values represent the best/the second-best result, respectively. The test dataset comprises 28 categories, and we report the averages for all while highlighting the top three.}
  \vspace{-0.1in}
  \setlength{\tabcolsep}{1mm}
  \resizebox{1.0\textwidth}{!}{
    \begin{tabular}{cc|cccc|cccc|cccc|rrrr}
    \toprule
    \multirow{2}[2]{*}{\textbf{\makecell[c]{Memory\\Setting}}} & \multirow{2}[2]{*}{\textbf{Model}} & \multicolumn{4}{c|}{\textbf{All}}      & \multicolumn{4}{c|}{\textbf{Home}} & \multicolumn{4}{c}{\textbf{Clothing}} & \multicolumn{4}{c}{\textbf{Tools}} \\
          &       & {R@100} & {R@10} & {N@100} & {N@10} & {R@100} & {R@10} & {N@100} & {N@10} & {R@100} & {R@10} & {N@100} & {N@10} & \multicolumn{1}{c}{{R@100}} & \multicolumn{1}{c}{{R@10}} & \multicolumn{1}{c}{{N@100}} & \multicolumn{1}{c}{{N@10}} \\
    \midrule
    \multirow{6}[2]{*}{\textbf{\textit{\makecell[c]{w/o\\Memory}}}} & GPT-4o & 0.226 & 0.092 & 0.086 & 0.059 & 0.227 & 0.057 & 0.090 & 0.033 & 0.100   & 0.030  & 0.036 & 0.022 & 0.247 & 0.097 & 0.096 & 0.065 \\
          & DeepSeek-R1 & 0.252 & 0.099 & 0.090  & 0.060  & 0.256 & 0.085 & 0.082 & 0.049 & 0.130  & 0.035 & 0.040  & 0.022 & 0.290 & 0.118 & 0.093 & 0.059 \\
          & Qwen2.5-3B & 0.255 & 0.096 & 0.097 & 0.065 & 0.261 & 0.081 & 0.081 & 0.043 & 0.120  & 0.035 & 0.036 & 0.020  & 0.290 & 0.108 & 0.114 & 0.078 \\
          & BLAIR-BASE & 0.227 & 0.072 & 0.070  & 0.040  & 0.213 & 0.062 & 0.056 & 0.025 & \textbf{0.135} & 0.035 & 0.037 & 0.018 &0.280 &0.086 &0.080 &0.042 \\
          & BLAIR-LARGE & 0.215 & 0.065 & 0.069 & 0.040  & 0.232 & 0.052 & 0.065 & 0.030  & 0.090  & 0.025 & 0.025 & 0.012 &0.312 &0.097 &0.104 &0.061 \\
          & Rec-R1 & 0.258 & \underline{0.111} & 0.099 & 0.071 & 0.265 & 0.085 & 0.086 & 0.047 & 0.126 & \underline{0.037} & 0.04  & 0.022 & 0.297 & 0.114 & 0.117 & 0.08 \\

    \midrule
    \multirow{6}[2]{*}{\textbf{\textit{\makecell[c]{w/ Naive\\Memory}}}} & GPT-4o & 0.258 & 0.109 & \underline{0.104} & 0.072 & \underline{0.278} & 0.081 & \underline{0.091} & 0.047 & 0.125 & 0.035 & 0.041 & 0.025 & 0.301 & \underline{0.119} & 0.110  & 0.079 \\
          & DeepSeek-R1 & \underline{0.260} & 0.106 & 0.100   & 0.067 & 0.275 & 0.085 & 0.090  & \underline{0.051} & 0.127 & 0.033 & \underline{0.043} & 0.026 &0.301 &0.118 &0.109 &0.074 \\
          & Qwen2.5-3B & 0.246 & 0.107 & 0.095 & 0.068 & 0.280  & \underline{0.088} & 0.084 & 0.049 & 0.105 & 0.035 & 0.037 & 0.026 &0.280 &0.108 &0.107 &0.073 \\
          & BLAIR-BASE & -     & -     & -     & - & -     & -     & -     & - & -     & -     & -     & - & - & - & - & - \\
          & BLAIR-LARGE & -     & -     & -     & - & -     & -     & -     & - & -     & -     & -     & - & - & - & - & - \\
          & Rec-R1 & \underline{0.260} & 0.108 & 0.097 & \underline{0.075} & 0.269 & 0.086 & 0.085 & 0.050 & 0.128 & 0.036 & 0.040 & \textbf{0.027} & 0.299 & 0.112  & \underline{0.119} & 0.085 \\

    \midrule
    \multirow{6}[2]{*}{\textbf{\textit{\makecell[c]{w/ Static\\Memory}}}} & GPT-4o & 0.252 & 0.098 & 0.095 & 0.065 & 0.237 & 0.076 & 0.071 & 0.041 & 0.105 & 0.030  & 0.037 & 0.022 & \underline{0.311} & 0.107 & 0.110  & 0.069 \\
          & DeepSeek-R1 & 0.249 & 0.089 & 0.087 & 0.057 & 0.232 & 0.076 & 0.072 & 0.044 & 0.125 & 0.025 & 0.038 & 0.020  & 0.301 & 0.086 & 0.092 & 0.050 \\
          & Qwen2.5-3B & 0.246 & 0.106 & 0.098 & 0.069 & 0.265 & 0.081 & 0.081 & 0.044 & 0.110  & 0.035 & 0.034 & 0.020  & 0.280 & 0.115 & 0.116 & \underline{0.086} \\
          & BLAIR-BASE & -     & -     & -     & - & -     & -     & -     & - & -     & -     & -     & - & - & - & - & - \\
          & BLAIR-LARGE & -     & -     & -     & - & -     & -     & -     & - & -     & -     & -     & - & - & - & - & - \\
          & Rec-R1 & 0.259 & 0.105 & 0.095 & 0.069 & 0.264 & 0.082 & 0.083 & 0.046 & 0.122 & 0.033 & 0.038 & 0.024 & 0.286 & 0.110  & 0.115 & 0.071 \\

    \midrule
    \multicolumn{2}{c|}{\textbf{Ours}} & \textbf{0.270 } & \textbf{0.122} &\textbf{ 0.113} & \textbf{0.084} & \textbf{0.284} & \textbf{0.090}  & \textbf{0.092} & \textbf{0.054} & \underline{0.130}  & \textbf{0.040 } & \textbf{0.045} & \textbf{0.027} & \textbf{0.333} & \textbf{0.129} & \textbf{0.132} & \textbf{0.091} \\

\multicolumn{2}{c|}{\cellcolor{gray!20}{Improvement}} & \cellcolor{gray!20}{+3.84\%}      &   \cellcolor{gray!20}{+9.91\%}    &  \cellcolor{gray!20}{+8.65\% }    &  \cellcolor{gray!20}{+12.00\%}     &  \cellcolor{gray!20}{+2.16\%}     &   \cellcolor{gray!20}{+2.27\%}    &    \cellcolor{gray!20}{+1.10\%}   &   \cellcolor{gray!20}{+5.88\%}    &  \cellcolor{gray!20}{-}     &    \cellcolor{gray!20}{+8.18\%}   & \cellcolor{gray!20}{+4.65\%}      & \cellcolor{gray!20}{+0.00\%}      &     \cellcolor{gray!20}{+7.07\%}  &   \cellcolor{gray!20}{+8.40\%}    &    \cellcolor{gray!20}{+10.92\%}   & \cellcolor{gray!20}{+5.81\%} \\ 
    \bottomrule
    \end{tabular}%
    }
  \label{tab:overall_performance}%
  \vspace{-0.05in}
\end{table*}%

We compare \ourmodel's recommendation accuracy with baselines, with results shown in Table~\ref{tab:overall_performance}. We make the following discussions:
\begin{itemize}[leftmargin=*, topsep=1pt, partopsep=0pt]
    \item \textbf{Performance Superiority of \ourmodel}. \ourmodel shows consistent advantages across different baseline categories. Compared to baselines without memories, our RAG-based memory mechanism improves performance by retrieving collaborative patterns. This advantage persists against baselines with naive memory and static summaries, demonstrating the effectiveness of synergizing memory with reasoning to enable dynamic preference exploration and contextual personalization. Compared to recommendation reasoning methods (\ie~Rec-R1 and BLAIR), \ourmodel extends reasoning to include iterative memory retrieval, enabling adaptive information gathering that evolves with the reasoning process rather than relying on static input, significantly enhancing reasoning depth and context awareness.
    
    \item \textbf{Incorporating Interaction Records for Baselines}. For all baselines, additionally incorporating naive memory (\ie~using raw interaction records) generally brings performance gains, validating the information gain of involving users' past behaviors in better understanding user queries. However, this improvement is marginal for baselines fine-tuned for recommendation (\ie~Rec-R1). This suggests that while LLM fine-tuning has unlocked models' recommendation potential, noise in raw interactions and limited reasoning depth prevent further improvements.
    
    \item \textbf{Static Summarization is Sometimes Harmful}. To handle noisy interaction records, fixed memory generates user summaries as external memories for the baseline methods. However, results show less improvement than naive memory and even cause performance degradation. This demonstrates the difficulty of building a global user summarization template, as the optimal solution may vary greatly for different users and item categories. Our dynamic memory retrieval mechanism with reasoning enhancement effectively addresses this concern and consistently delivers substantial performance gains across all scenarios.
\end{itemize}

\subsection{Ablation Study (RQ2)}
\begin{figure}[t]
    \centering
    \vspace{-0.01in}
    \includegraphics[width=1\linewidth]{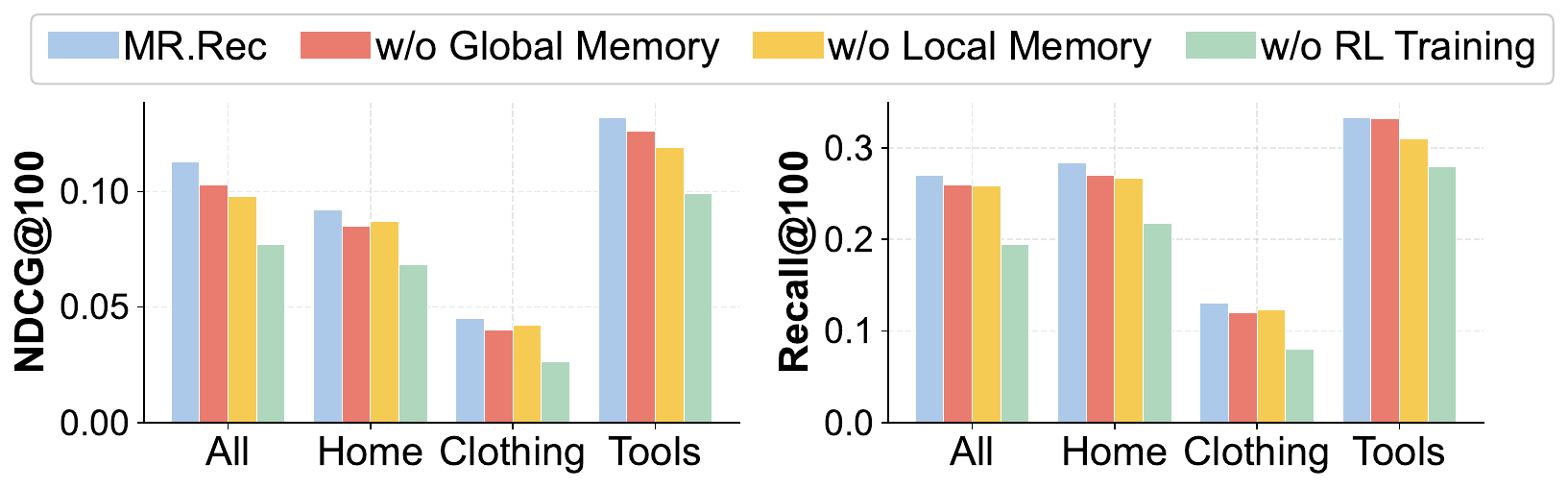}
    \vspace{-0.25in}
    \caption{Ablation of different components of \ourmodel.}
    \label{fig:ablation}
    \vspace{-0.25in}
\end{figure}
\subsubsection{\bf Effect of Key Components}
We first study the impact of different technical components of \ourmodel. The evaluation results are presented in Figure~\ref{fig:ablation}. We study the following components:
\begin{itemize}[leftmargin=*,itemsep=0pt, parsep=0pt, topsep=0pt, partopsep=0pt]
    \item \textbf{w/o Local/Global Memory}. The results demonstrate that removing either local or global memory mechanisms causes significant performance degradation, with varying impact across different datasets. This validates the effectiveness of our memory indexing method in capturing beneficial interaction patterns and our retrieval method in effectively accessing this information.

    \item \textbf{w/o RL Tuning}. Compared to removing memory mechanisms, directly using the base LLM without RL tuning causes larger performance damage. This is because RL tuning not only improves the base model's performance, but also significantly impacts the retrieval loop effectiveness. Without our tuning, small-parameter LLMs such as the 3B model struggle to determine when and how to leverage memory and perform reasoning. This highlights the critical role of RL tuning for \ourmodel.
\end{itemize}

\vspace{-0.05in}
\subsubsection{\bf Effect of Local Memory Components}
As shown in Table~\ref{tab:ablation_local_memory}, we further investigate the impact of different components of user-specific local memory on model performance. The results reveal that the absence of the entire local memory mechanism leads to the weakest performance, whereas incorporating any single component (\textbf{B}ehavior Records, \textbf{P}reference Patterns, or \textbf{U}ser Profile) individually yields clear improvements. The best results are achieved when all three components are combined, suggesting a complementary effect that more comprehensively captures user preferences relevant to the current query.
\begin{table}[t]
  \centering
  \caption{Ablation of user-specific local memory components.}
  \small
  \setlength{\tabcolsep}{2.4mm}
  \vspace{-0.1in}
  {
    \begin{tabular}{lrrrr}
    \toprule
          \textbf{Variant} & \multicolumn{1}{c}{{\textbf{R@100}}} & \multicolumn{1}{c}{{\textbf{R@10}}} & \multicolumn{1}{c}{{\textbf{N@100}}} & \multicolumn{1}{c}{{\textbf{N@10}}} \\
    \midrule
    w/o Local Memory  &   0.258    &   0.113    &    0.098   &  0.069 \\
    w/ \textbf{B}ehavior Records &   0.268$\uparrow$    &  0.117$\uparrow$     &   0.109$\uparrow$    & 0.079$\uparrow$ \\
    w/ \textbf{P}reference Patterns &  \textbf{0.272}$\uparrow$    &   0.116$\uparrow$    & 0.110$\uparrow$      & 0.079$\uparrow$ \\
    w/ \textbf{U}ser Profile &    0.269$\uparrow$   &  0.113    &  \textbf{0.113}$\uparrow$     & 0.080$\uparrow$ \\
    w/ \textbf{B}+\textbf{P}+\textbf{U} &  0.270$\uparrow$     &  \textbf{0.122}$\uparrow$     &  \textbf{0.113}$\uparrow$     &  0.084$\uparrow$ \\
    \bottomrule
    \end{tabular}%
    }
  \label{tab:ablation_local_memory}%
  \vspace{-0.225in}
\end{table}%

\vspace{-0.05in}
\subsection{Memory Effectiveness Study (RQ3)}
To further investigate the effectiveness of memory in the recommendation process, we analyzed the retrieved memory and designed two metrics: \textbf{Memory-to-Profile Contribution (MPC) }and \textbf{Memory-to-Recommendation Contribution (MRC)}. MPC measures whether the retrieved memory aids in generating the ideal item profile, while MRC evaluates whether it contributes to the final correct recommendation.

To ensure a comprehensive and objective evaluation, we employed two assessment approaches: a \textbf{heuristic method}, checking keyword overlaps between retrieved memory and profiles or ground-truth items, and an \textbf{LLM-judged method}, where an LLM determines whether the memory is helpful.

As shown in Table~\ref{tab:memory_analysis}, both Global and Local Memory are valuable in the recommendation process. For MPC, they both achieve high scores, indicating that retrieved memory significantly aids the generation of ideal item profiles. This demonstrates that \ourmodel effectively leverages memory to provide final recommendation. For MRC, Global Memory scores relatively high under the heuristic method but somewhat lower under LLM judgment, reflecting that it often overlaps with ground-truth items at the surface level, while semantic contribution is more nuanced. By contrast, Local Memory shows modest heuristic scores but substantially higher LLM-judged scores. This indicates that, despite limited surface-level overlap, local memory provides more semantically meaningful support for recommendations. This effectiveness arises from our approach: instead of one-shot retrieval based solely on query similarity, the LLM is trained to reason about which memory entries are genuinely useful, enabling more precise and context-aware recommendations.

\begin{table}[t]
  \centering
   \vspace{-0.15in}
  \caption{Contribution of retrieved memory.}
  \small
  \setlength{\tabcolsep}{1.6mm}
  \vspace{-0.15in}
  \resizebox{0.45\textwidth}{!}{
    \begin{tabular}{lccccc}
    \toprule
    \multicolumn{1}{c}{\multirow{2}[2]{*}{\textbf{Memory Type}}} & \multicolumn{2}{c}{\textbf{MPC}} &&\multicolumn{2}{c}{\textbf{MRC}} \\
    \cline{2-3}
    \cline{5-6}
    & Heuristic & LLM-Judged & & Heuristic & LLM-Judged \\
    \midrule
    Global Memory &  0.9605     & 0.9172    &  &   0.9012    & 0.5375 \\
    Local Memory & 0.9368 & 0.9511 & & 0.4743 & 0.7036 \\
    \bottomrule
    \end{tabular}%
    }
    \vspace{-0.2in}
  \label{tab:memory_analysis}%
\end{table}%

\subsection{Efficiency Study (RQ4)}
\begin{table}[b]
  \centering
  \vspace{-2em}
  \caption{Cost and Efficiency for Memory Indexing and Retrieval in the \ourmodel Framework}
  \vspace{-1.3em}
  \small
  \setlength{\tabcolsep}{2.2mm}
  \resizebox{0.49\textwidth}{!}{
    \begin{tabular}{lcccc}
    \toprule
    \rowcolor{gray!20}
    \multicolumn{5}{c}{\textbf{Statistics and Cost of Memory Indexing}} \\
    \midrule
    & \multicolumn{2}{c}{\textbf{Local Memory}} & \multicolumn{2}{c}{\textbf{Global Memory}} \\
    \midrule
    Source Users & \multicolumn{2}{c}{3,000} & \multicolumn{2}{c}{157} \\
    Input Tokens (M) & \multicolumn{2}{c}{385.21} & \multicolumn{2}{c}{0.48} \\
    Output Tokens (M) & \multicolumn{2}{c}{42.00} & \multicolumn{2}{c}{0.06} \\
    Memory Entries & \multicolumn{2}{c}{73,078} & \multicolumn{2}{c}{1,970} \\
    Total Time (s) & \multicolumn{2}{c}{2,435} & \multicolumn{2}{c}{121} \\
    Per-entry Time (s) & \multicolumn{2}{c}{0.07} & \multicolumn{2}{c}{0.06} \\
    API Cost (GPT-4o-mini) & \multicolumn{2}{c}{54.09} & \multicolumn{2}{c}{0.07} \\
    Per-entry Cost & \multicolumn{2}{c}{$7.4 \times 10^{-4}$} & \multicolumn{2}{c}{$3.6 \times 10^{-5}$} \\
    \midrule
    \rowcolor{gray!20}
    \multicolumn{5}{c}{\textbf{Token Efficiency of Memory Retrieval}} \\
    \midrule
    \multicolumn{5}{c}{
      \begin{tabular}{cccc}
        Memory Setting & \makecell[c]{Avg. Memory\\Tokens} & \makecell[c]{R@100\\(GPT-4o)} & \makecell[c]{Efficiency\\(R@100/100 tokens)} \\
        \midrule
        Recent 10 interactions & 283.51 & 0.261 & 0.092 \\
        Static user profile & 492.7 & 0.261 & 0.053 \\
        \ourmodel & 95.43 & 0.285 & 0.299 \\
      \end{tabular}
    } \\
    \bottomrule
    \end{tabular}
  }
  \vspace{-1em}
  \label{tab:memory_efficiency_all}
\end{table}

This experiment studies the costs and efficiency of our memory indexing and retrieval mechanisms. We built user-specific local memory for 3,000 users and sampled 157 users for cross-user global memory, creating 73,078 local and 1,970 global memory entries. The statistics and costs are shown in Table~\ref{tab:memory_efficiency_all}.

\begin{itemize}[leftmargin=*,itemsep=0pt, parsep=0pt, topsep=0pt, partopsep=0pt]
    \item \textbf{Time and Cost Efficiency of Indexing}. In our proposed \ourmodel, indexing time per entry is 0.07s for local and 0.06s for global memory, with API costs of \$54 for 3,000 users (\$0.018/user) and \$0.07 for global memory. Global memory uses only 0.48M input and 0.06M output tokens, roughly 1/1000 of local memory, showing high indexing efficiency.
    \item \textbf{Token Efficiency of Retrieval}. Table~\ref{tab:memory_efficiency_all} shows our method achieves highest R@100 with 95.43 tokens, yielding an efficiency of 0.299 (R@100/100 tokens), outperforming baselines that utilize recent interactions or static profiles as memory. By retrieving the most relevant memory entries for recommendation, our method introduces less noise into the context, leading to higher recall and more efficient reasoning.

\end{itemize}

\subsection{Sensitivity Study (RQ5)}
This section studies the impact of different configurations of our \ourmodel, including LLM backbones and different retrieval settings.

\subsubsection{\bf Impact of LLM Backbones}
\begin{figure}[t]
    \centering
    \vspace{-0.2in}
    \includegraphics[width=1\linewidth]{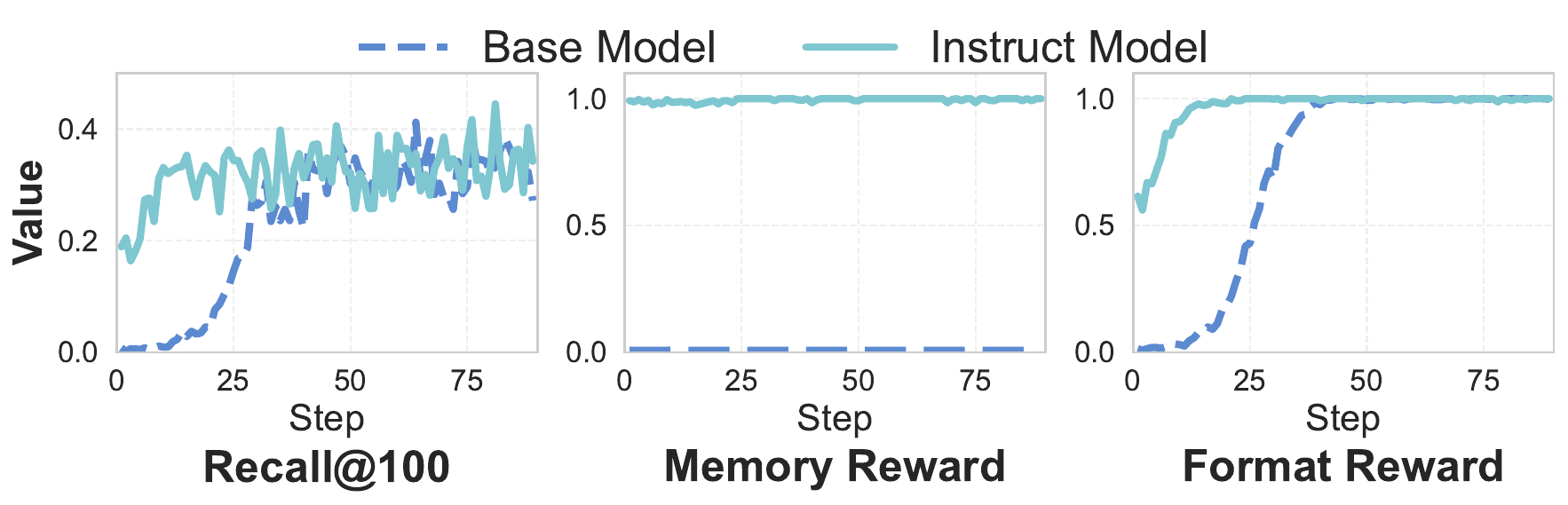}
    \vspace{-0.3in}
    \caption{Reward trajectories of \ourmodel using the base and instruct models of Qwen-2.5-3B, during our RL tuning.}
    \label{fig:backbone}
    \vspace{-0.2in}
\end{figure}

To study the impact of different LLM backbones, we compare the default backbone of \ourmodel, Qwen-2.5-3B-Instruct, with its base model Qwen-2.5-3B-Base. The former has undergone prior instruction tuning and reinforcement learning from human feedback (RLHF) to strengthen its instruction-following capabilities, while the latter has not. Both \ourmodel variants using these backbone LLMs are tuned using our RL method, and their reward trajectories are recorded and presented in Figure~\ref{fig:backbone}.

The results demonstrate strong instruction-following ability of the Instruct model from the start, correctly invoking memory retrieval (Memory reward) and generating properly formatted outputs (Format reward). In contrast, the Base model initially struggles to follow instructions, with format reward remaining near zero. During training, the Base model gradually learns to produce properly formatted outputs and shows improvements in recommendation performance, as reflected by Recall@100. However, it consistently fails to learn memory retrieval, as indicated by near-zero memory reward, instead relying solely on direct recommendations.

\subsubsection{\bf Impact of Base Retrievers}
We compared using three different base retrieval algorithms for our reasoning-enhanced retrieval process, including Qwen3-Embedding-0.6B~\cite{zhang2025qwen3}, BGE-M3~\cite{chen2024bge}, and BM25~\cite{robertson2009probabilistic}. Among these, Qwen3-Embedding-0.6B supports customized instructions during retrieval, BGE-M3 is a pre-trained deep learning model, while BM25 is a traditional lexical matching method that requires no pre-training.

\begin{figure}
    \centering
    \includegraphics[width=0.86\linewidth]{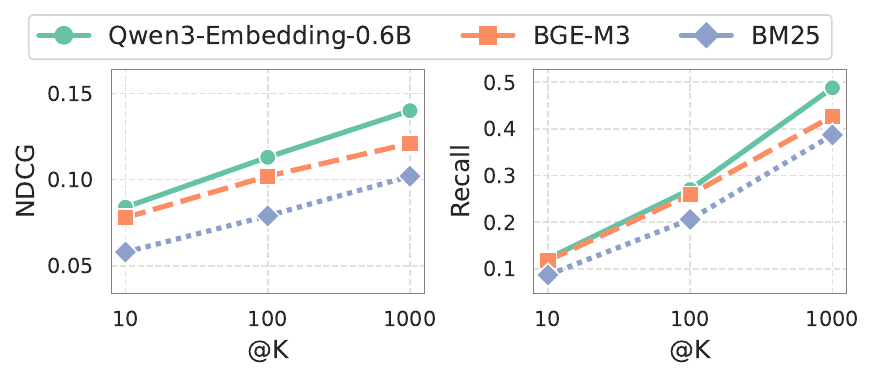}
    \vspace{-0.2in}
    \caption{Effect of using different retrievers for \ourmodel.}
    \label{fig:different_retrievers}
    \vspace{-0.1in}
\end{figure}

The experimental results shown in Figure~\ref{fig:different_retrievers} demonstrate that Qwen3-Embedding-0.6B achieves the best performance, while BM25, which lacks specialized training for dense retrieval, shows a significant performance gap compared to the embedding-based models. The performance gaps between models become more pronounced as the value of k increases in NDCG and Recall@k metrics, indicating that the advantages of instruction-tuned embedding models are more evident when retrieving larger sets of relevant items.

\subsubsection{\bf Impact of Retrieved Memory Size.}
\begin{figure}[t]
    \centering
    \vspace{-0.1in}
    \includegraphics[width=1\linewidth]{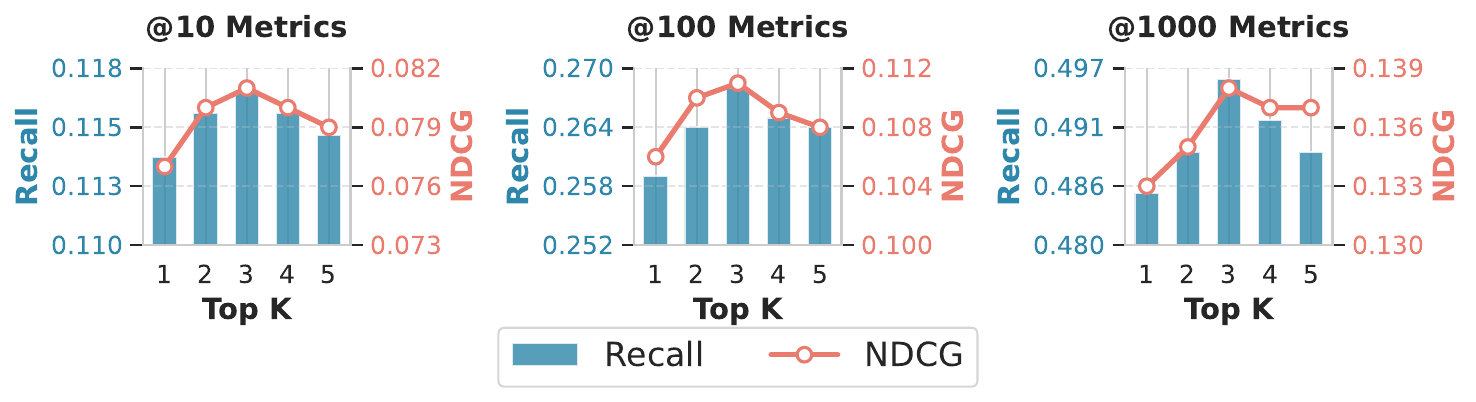}
    \vspace{-0.25in}
    \caption{Effect of different top-k of memory retrieval.}
    \label{fig:retrieved_topk}
    \vspace{-0.2in}
\end{figure}

We further analyze the impact of retrieving different numbers of top-k relevant memory entries on model performance, with results shown in Figure~\ref{fig:retrieved_topk}. The results reveal a trade-off pattern: smaller k values may miss useful memory information, while larger k values may introduce noise and extend the context window, potentially degrading LLM performance. Based on these findings, we select k=3 as the optimal balance, which consistently achieves strong performance.

\subsection{Case Study}
\begin{figure}[t]
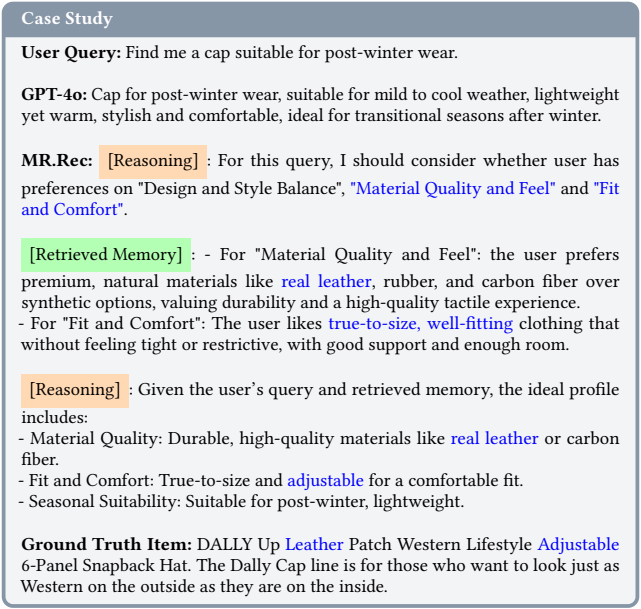
{
\footnotesize
\begin{cmt*}{Case Study}{}
\textbf{User Query:} Find me a cap suitable for post-winter wear.
\\

\textbf{GPT-4o:} Cap for post-winter wear, suitable for mild to cool weather, lightweight yet warm, stylish and comfortable, ideal for transitional seasons after winter.
\\

\textbf{\ourmodel:} \colorbox{orange!30}{[Reasoning]}: For this query, I should consider whether user has preferences on "Design and Style Balance", \textcolor{blue}{"Material Quality and Feel"} and \textcolor{blue}{"Fit and Comfort"}.
\\

\colorbox{green!30}{[Retrieved Memory]}:
- For "Material Quality and Feel": the user prefers premium, natural materials like \textcolor{blue}{real leather}, rubber, and carbon fiber over synthetic options, valuing durability and a high-quality tactile experience.
\\
- For "Fit and Comfort": The user likes \textcolor{blue}{true-to-size, well-fitting} clothing that without feeling tight or restrictive, with good support and enough room. 
\\

\colorbox{orange!30}{[Reasoning]}: Given the user’s query and retrieved memory, the ideal profile includes:\\
- Material Quality: Durable, high-quality materials like \textcolor{blue}{real leather} or carbon fiber.\\
- Fit and Comfort: True-to-size and \textcolor{blue}{adjustable} for a comfortable fit. \\
- Seasonal Suitability: Suitable for post-winter, lightweight.
\\

\textbf{Ground Truth Item:}  DALLY Up \textcolor{blue}{Leather} Patch Western Lifestyle \textcolor{blue}{Adjustable} 6-Panel Snapback Hat. The Dally Cap line is for those who want to look just as Western on the outside as they are on the inside. 

\end{cmt*}
}
\vspace{-0.2in}
\caption{Case study demonstrating \ourmodel's memory-synergized reasoning for personalized cap recommendation.}
\label{fig:case_study}
\vspace{-0.2in}
\end{figure}

We present a case study in Figure~\ref{fig:case_study}, illustrating how \ourmodel integrates memory and reasoning for recommendation. When a user queries for "a cap suitable for post-winter wear," the baseline GPT-4o generates a generic response focused on seasonal suitability. In contrast, \ourmodel first reasons through implicit preference dimensions like "Material Quality and Feel" and "Fit and Comfort." It then retrieves relevant memory segments reflecting the user's preference for premium materials like real leather and adjustable items. The model generates an ideal item profile combining these preferences with seasonal needs. The final recommendation strongly aligns with the ground truth item, a leather patch adjustable snapback hat, with key attributes (leather, adjustable) appearing in both the generated profile and actual item, demonstrating how our reasoning-enhanced memory retrieval successfully captures user-specific preferences that would be missed by generic baselines.

\vspace{-0.1in}
\subsection{Related Work}
\subsubsection{\bf Memory Modeling in Recommender Systems}
Memory modeling is essential for capturing dynamic user preferences in RSs. Traditional sequential models, such as SASRec~\cite{kang2018self}, implicitly encode user history but often struggle with long sequences and lack explicit interpretability. The emergence of LLMs has introduced explicit memory mechanisms, enabling more sophisticated and interpretable representations of user behavior. A prominent approach involves utilizing user interaction history as memory~\cite{wang2025knowledge,huang2025towards,zhang2025llm}, with methods like AutoMR~\cite{wang2025leveraging} employing learned retrievers to intelligently select relevant historical segments. In contrast, MARM~\cite{lv2024marm} and LMN~\cite{lu2025large} enhance efficiency by caching computations and utilizing product quantization to handle long sequences and large memory banks. An alternative approach leverages LLM-generated user profiles, which synthesize extensive interaction histories into textual summaries~\cite{zhang2024guided}. MAP~\cite{chen2025memory} constructs user profiles as tables of historical ratings and reviews, but this flat memory structure struggles with noisy and multi-granularity preferences. MemoCRS~\cite{xi2024memocrs} introduces an entity-based user memory and a shared general memory for multi-turn dialogues. Nevertheless, it fails to effectively integrate memory with reasoning, resulting in a superficial reliance on similarity-based retrieval, which neglects deeper semantic relationships within the memory and limits the generation of contextually relevant recommendations.
In comparison, \ourmodel develops hierarchical memory indexing and reasoning-enhanced retrieval, better filtering noisy information while achieving efficiency.

\subsubsection{\bf LLM Reasoning for Recommendation}
LLMs have transformed RSs by introducing advanced reasoning beyond traditional collaborative filtering. Key developments include Chain-of-Thought (CoT)~\cite{wei2022chain} reasoning, which allows LLMs to generate step-by-step logical recommendations. For instance, HetGCoT-Rec~\cite{jia2025hetgcot} integrates heterogeneous graphs with CoT for journal recommendations, and Li et al.~\cite{li2024identify} use CoT to enhance user intention inference and contextual awareness. However, prompting-based CoT methods remain limited in complex recommendation tasks. To address these limitations, recent studies have shifted toward training LLMs explicitly for reasoning rather than relying solely on prompting. A representative approach for enhancing LLM reasoning is GRPO~\cite{guo2025deepseek}. Reason-to-Recommend~\cite{zhao2025reason} employs GRPO to strengthen "Interaction-of-Thought" reasoning, enabling emergent planning behaviors in challenging recommendation tasks. Similarly, LatentR~\cite{zhang2025reinforced} adopts a modified GRPO framework to optimize latent reasoning trajectories, avoiding explicit text generation while preserving reasoning capacity.
Our work extends LLM finetuning beyond CoT reasoning to include memory-synergized reasoning.

\section{Conclusion}
In this paper, we introduced \ourmodel, a framework that unifies memory and reasoning to advance LLM-based recommendation assistants. By combining a RAG-based external memory mechanism with reinforcement learning, our approach enables the LLM to actively explore and select useful preferences and experiences while performing multi-step reasoning over retrieved information to generate accurate recommendations. Extensive experiments on multiple benchmark datasets demonstrate that \ourmodel consistently outperforms state-of-the-art baselines in both personalization and reasoning capability. We believe \ourmodel takes a meaningful step toward intelligent and user-centered recommendation assistants.


\clearpage
\bibliographystyle{plain}
\balance
\bibliography{ref}

\appendix
\newpage
\noindent\textbf{\LARGE APPENDIX}
\section{Dataset}
\label{app:dataset}

We derive our experimental data from the Amazon-C4 corpus, which contains user queries synthesized by ChatGPT from product review text. These synthetic queries often include fine-grained, product-level details (e.g., brand names, model identifiers, and exact technical specifications) that are uncommon in real-world user utterances. In practice, users tend to give concise or underspecified requests—such as “recommend a budget laptop for college” or “gift ideas for my father who likes hiking”—and expect the recommender to infer context from prior interactions or stored user memories. To better emulate these memory-driven recommendation scenarios, we transform Amazon-C4 so that queries resemble realistic, high-level requests. Concretely, we (i) remove or mask overly specific attributes (brand names, exact model identifiers, serial numbers, and verbatim review excerpts) and (ii) preserve the user’s core intent and constraints (product category, coarse price band, primary use case). To achieve this, we prompt GPT-o3-mini to produce concise, naturalistic rewrites that retain the original intent and constraints while omitting gratuitous detail. The prompt used for simplification and representative before/after examples are provided below.
\begin{figure}[t]
    \centering
 \begin{cmt*}{Prompt Template for Pre-processing Queries}{}
    \textbf{System Prompt:} Rewrite the "Original Query" into a single, casual, conversational question.
    Rules: \\
    - Keep: 1 core item + 1 core use case (e.g., "a laptop for gaming").\\
    - Delete: All other details (including brand, specs, price, personal preferences, etc.).\\
    - Tone: Natural and conversational.\\
    - Format: Must be a single sentence.\\
\\
    \textbf{Original Query}: \textit{I'm in need of an external Blu-ray drive that actually works well with reading Blu-rays. The one I currently own has a poor ability to read discs, requiring multiple tries and hoping for a successful read. I want a Blu-ray drive that is amazing by comparison, where I can simply insert a disc and it starts playing right away. Skipping around on the disc should be smooth and there shouldn't be any endless seeking from the drive. I came across one drive that shows up as a Pioneer BDR-TD03 in the properties, and based on its performance, I'm really satisfied with its quality. The name "Dainty" might be unusual for a disk drive, but it doesn't affect its functionality, which is fantastic.}\\

    \textbf{Rewritten Query}: \textit{Hey, can you recommend an external Blu-ray drive that reads discs reliably and plays them smoothly right from the start?}\\
\end{cmt*}
\end{figure}

\begin{table}[t]
  \centering
  \caption{Dataset Statistics}
  \vspace{-1em}
    \begin{tabular}{cc}
    \toprule
    Num. Queries & 6096 \\
    Num. Users & 6039 \\
    Num. Categories & 28 \\
    Num. Candidate Items & 1,058,417 \\
    Num. User-Item Interactions & 577,376 \\
    \bottomrule
    \end{tabular}%
    \vspace{-1em}
  \label{tab:dataset}%
\end{table}%

Due to the limited availability of API resources, we randomly sampled 6,096 queries from the Amazon-C4 dataset and applied the aforementioned preprocessing procedure to this subset. The resulting dataset retains the essential characteristics necessary for our experiments while remaining computationally manageable. A detailed summary of the processed dataset is provided in Table~\ref{tab:dataset}.

\section{Implementation Details}
\label{app:imp}

\subsection{Environment} Our experiments were conducted using 4 x NVIDIA A800(80GB) GPUs. We employed PyTorch 2.6.0 as the deep learning framework, Transformers 4.48.0 for model implementation, and the GRPO implementation provided by VeRL. For inference acceleration, we utilized SgLang 0.4.6.
\subsection{Prompts for LLM-based Baselines} With the exception of BLAIR, all other baselines are LLM-based. Following the Rec-R1 framework, we prompt the LLM to generate an ideal item profile corresponding to a given user query. Specifically, we employ the following prompt template: \textit{"You are an expert in generating queries for dense retrieval. Given a customer
query, your task is to retain the original query while expanding it with
additional semantically relevant information, retrieve the most relevant
products, ensuring they best meet customer needs. If no useful expansion is
needed, return the original query as is.
\# Below is the product search query:\{query\}"}
\subsection{Details for Our Method} For both memory retrieval and item retrieval, we adopt the Qwen3-Embedding-0.6B model as the retrieval encoder. For reinforcement learning with LLMs, the learning rate is set to 1e−5 with a batch size of 256. During trajectory generation, we configure the group size (G = 5) and use a rollout temperature of 1.0.

\section{Prompts Used in Our Method}
The prompts employed in our method fall into two primary categories: \textbf{Memory Indexing} and \textbf{LLM-based Recommendation}. 
\subsection{Prompts for Memory Indexing}
\subsubsection{Cross-User Global Memory}
\begin{figure*}[h!]
    \centering
    \begin{cmt*}{Prompt for Cross-User Global Memory (A)}{}
    You are given the following information for a recommendation task:

-Recommendation scenario: \textcolor{blue}{\{scenario\}}

-User query: \textcolor{blue}{\{query\}}

-Ground truth item (the item finally purchased): 
\textcolor{blue}{\{ground truth item metadata\}}

-Negative items (similar items not chosen by the user): 
\textcolor{blue}{\{negative items\}}
\\

Your task:

1. Analyze the user query, the purchased item, and the unchosen items.

2. Identify additional aspects or dimensions that the user might have implicitly considered beyond what is explicitly stated in the query. These aspects should support more personalized recommendations.

\* At the category level: What general personalization factors might users care about beyond the query, such as overall budget, preferred brands, aesthetics, or usage context? Do not mention specific product or features in this level.

\* At the subcategory level (e.g., the relevant subcategory of ground truth item): What specific personalization factors might be relevant, especially when comparing the ground truth item with the negative items (e.g., screen size, smart features, energy efficiency, style)?

3. Focus on uncovering subtle or implicit preferences (e.g., color, price sensitivity, design style, brand affinity, feature trade-offs) that can be inferred by comparing the ground truth item with the negative items.
\\

Output format (structured):
\begin{verbatim}
{{
    "category_level_personalization_aspects": [
        {
            "category":  "category_name",
             "aspect":  "aspect_name",
             "description":  "description"
        },
        ...]
    ,
    "subcategory_level_personalization_aspects": [
        {
             "subcategory":  "subcategory_name",
             "aspect":  "aspect_name",
             "description":  "description "
        },
        ...
    ]
}}
\end{verbatim}
\end{cmt*}
    \vspace{-0.2in}
\end{figure*}
To construct the Cross-User Global Memory, we sample interaction histories from multiple users across different recommendation scenarios (e.g., purchasing products from various categories). These histories include user queries, selected items, and their associated metadata. The LLM is then prompted to analyze each scenario and infer additional aspects that users might consider beyond what is explicitly mentioned in the query, thereby extracting richer recommendation experiences. To encourage the LLM to better reason about these additional aspects, we also sample items that are similar to the ground-truth purchases but were not selected by the user as negative items. The prompt used in this process is as above.

Using the prompt described above, the LLM analyzes each user’s interaction data to extract recommendation knowledge specific to the corresponding scenario. To capture generalizable cross-user global memory, we sample interactions from multiple users and prompt the LLM to aggregate the extracted knowledge, producing a concise and coherent summary that represents common recommendation considerations within the scenario. The prompt used for this process is presented below:

\begin{figure*}[h!]
    \centering
    \begin{cmt*}{Prompt for Cross-User Global Memory (B)}{}
 You are given some aspects that multiple users consider important for a category.

Category: \textcolor{blue}{\{category\}}

Current aggregated aspects:
\textcolor{blue}{\{aspects\}}
\\

Task:
- Merge semantically duplicate or highly overlapping aspects into unified canonical names.
- Normalize naming style (use consistent Title Case English or best canonical phrasing).
- Consolidate and de-duplicate descriptions under each merged aspect; keep key insights, remove redundancy.
- These are shared aspects of multiple users, so do not use any specific product/parameter/user in the description.
\\

Output strictly as a compact JSON object mapping aspect\_name to list of descriptions, e.g.:

\begin{verbatim}
{{
    "aspect_name": ["description", ...],
    ...
}}
\end{verbatim}
\end{cmt*}
    
\end{figure*}

\subsubsection{User-specific Local Memory}
Our proposed User-specific Local Memory is organized into three hierarchical levels: Behavior Records, Preference Patterns, and User Profile. These levels capture user information at progressively finer granularity, ranging from abundant and potentially noisy interaction data to highly abstracted and fine-grained summaries. Specifically, Preference Patterns are derived from*Behavior Records, and the User Profile represents a further distillation and consolidation of these preference patterns, providing a succinct yet informative summary of the user’s interests and tendencies.
\begin{figure}[h!]
    \centering
    \begin{cmt*}{Prompt for Extracting Preference Patterns}{}
 You are given the following information about a user’s purchase history in a specific category:
\\

- Category: \textcolor{blue}{\{category\}}
- User’s purchased items in this category, each with its metadata and the user’s review:
\textcolor{blue}{\{reviews\}}
\\

Your task:
Analyze the items (including their metadata and the user’s reviews) and summarize the user’s preferences in this category. 
The summary should capture consistent patterns across items and reviews (e.g., favored brands, preferred price range, styles, features, quality expectations). 
Be as detailed as possible, but do not fabricate information that is not supported by the input. 
Directly output the preference summary (string) below:

\end{cmt*}
    
\end{figure}

For a given user, we collect their interaction history within a specific recommendation scenario and prompt the LLM to analyze these interactions to identify Preference Patterns. This process enables the extraction of recurring behaviors, interests, and implicit constraints that characterize the user’s preferences within the scenario. The prompt used for this analysis is as follows:

After generating Preference Patterns for a user across different scenarios, we prompt the LLM to reason over these patterns to infer the user’s potential User Profile. This step consolidates scenario-specific preferences into a coherent and high-level summary of the user’s interests and characteristics. The prompt used for this inference process is as follows:

\begin{figure}[h!]
    \centering
    \begin{cmt*}{Prompt for Generating User Profile}{}
 You are given the following information about a user’s preferences across different categories or aspects:

 \textcolor{blue}{\{preference patterns\}}
\\

Your task:
Based on these preferences, infer the user’s overall profile.
The profile should summarize general traits, patterns, and tendencies that can be reasonably inferred from the given preferences (e.g., spending habits, brand inclination, style choices, feature priorities, quality expectations, lifestyle hints). 
Do not fabricate information that is not supported by the input.
Directly output the profile (string) below:

\end{cmt*}
    
\end{figure}

\subsection{Prompts for LLM-based Recommendation}
In this study, we develop an LLM-based recommendation assistant that integrates memory and reasoning. To enable the LLM to autonomously leverage user-specific local memory guided by cross-user global memory, we employ the following prompt, which instructs the model to generate keywords for global memory retrieval. The LLM then uses the retrieved memory to perform reasoning and produce personalized recommendations.
\begin{figure}[h!]
    \centering
    \begin{cmt*}{Prompt for LLM-based Recommendation}{}
 You are a recommendation expert. Your task is to generate an ideal item profile based on a user's query and memories.
When given a user query, you should analyze the user's query and identify additional factors that might refine the recommendation.  \textcolor{blue}{\{Retrieved Cross-User Global Memory Based on Query\}}
\\

You should retrieve current user's memories that related to these aspects via `memory\_retrieval\_tool`. 
After retrieving the memories, combine them with the user's query, think step by step to reason about an ideal item profile for item retrieval. This profile should be as detailed as possible, do not miss any important keywords, otherwise relevant items may be missed. Your final output should be structured as follows:
\begin{verbatim}
\boxed{"ideal_item_profile": str, 
"useful_memory_ids": List[str]}
\end{verbatim}
Remember: First, retrieve the relevant memories, then reason about detailed ideal item profile based on the query and the memories you retrieved.

\end{cmt*}
\vspace{-0.2in}
\end{figure}

\end{CJK}
\end{document}